# Cultivating the Garden of Eden

Randall D. Beer
Cognitive Science Program
Informatics Dept.
Luddy School of Informatics, Computing and Engineering
Indiana University

**Abstract**

Garden of Eden (GOE) states in cellular automata are grid configurations which have no precursors, that is, they can only occur as initial conditions. Finding individual configurations that minimize or maximize some criterion of interest (e.g., grid size, density, etc.) has been a popular sport in recreational mathematics, but systematic studies of the set of GOEs for a cellular automaton have been rare. This paper presents the current results of an ongoing computational study of GOE configurations in Conway's Game of Life (GoL) cellular automaton. Specifically, we describe the current status of a map of the layout of GOEs and nonGOEs in density/size space, characterize how the density-dependent structure of the number of precursors varies with increasing grid size as we approach the point where GOEs begin to occur, provide a catalog of all known GOE configurations up to a grid size of 11 × 11, and initiate a study of the structure of the network of constraints that characterize GOE vs. nonGOE configurations.

Please address all correspondence to:

Randall D. Beer                                  Phone: (812) 856-0873
Cognitive Science Program                        Fax: (812) 855-1086
3052 Geology Building                            Email: rdbeer@indiana.edu
Indiana University                               URL: https://rdbeer.pages.iu.edu/
Bloomington, IN 47405

# 1. Introduction

The Game of Life (GoL) is a well-known cellular automaton that was invented by John Conway and originally popularized by Martin Gardner in the pages of *Scientific American* in 1970 (Gardner, 1970; Berlekamp, Conway, & Guey, 1982; Adamatzky, 2010; Johnston & Greene, 2021). In GoL, each cell in a square two-dimensional grid can be either ON or OFF. A cell changes state according to a simple rule based on its own state and the states of the eight cells immediately surrounding it. Continuing interest in this cellular automaton derives from the remarkable complexity that arises from the temporal evolution of this simple rule. Indeed, it is well-known that the Game of Life is computationally universal.

Of the many interesting questions raised by the Game of Life, the Garden of Eden problem has been one of the most enduring. A Garden of Eden (GOE) configuration is one that has no precursors, that is, that can only occur as an initial state in the time evolution of a cellular automaton. The first GOE configuration in the Game of Life, a $33 \times 9$ pattern containing 226 ON cells, was discovered in 1971 by Roger Banks and verified by Don Woods (Gardner, 1983:230,248; Poundstone, 1985:49). Since that time, a small but devoted group of enthusiasts has searched the space of Game of Life configurations, informally competing to find ever smaller or less dense GOEs (LifeWiki, 2022; Flammenkamp, 2009).

A significant disadvantage of the focus on breaking records is that it tends to emphasize the characteristics of individual GOEs. This is unfortunate because GOEs are much more than just an entertaining challenge in recreational mathematics. Because GOEs serve as the basin boundaries of distinct global attractors, they play a central role in organizing the global dynamics of a cellular automaton. Accordingly, it would be useful to have a more systematic understanding of the properties of the ensemble of GOEs as a whole.

Work along these lines was recently initiated by Hartman et al. (2013), who generated all square GOE configurations in the Game of Life respecting certain symmetries up to a grid size of $14 \times 14$. They found that no GOEs exist below a grid size of $6 \times 6$ and that no GOEs satisfying certain symmetries exist below a grid size of $10 \times 10$; the question of whether or not non-symmetric GOEs exist at intermediate grid sizes remains open. In addition, they found that the occurrence of GOEs was strongly dependent on the 1-density of configurations. Specifically, the number of symmetric GOEs was found to be normally-distributed with respect to density.

The goal of this paper is to extend this analysis in four ways. First, we present the current status of a comprehensive map of the layout of GOEs and nonGOEs in density/size space. Second, we characterize how the density-dependent structure of the number of precursors varies with increasing grid size as we approach the point where GOEs begin to occur. Third, we provide and analyze a catalog of all known GOE configurations up to a grid size of $11 \times 11$. Fourth, we initiate a study of the structure of the network of constraints that characterize GOE vs. nonGOE configurations.

## 2. Preliminaries

Conway's Game of Life is a binary outer totalistic cellular automaton on a two-dimensional rectangular grid (Gardner, 1970; Berlekamp, Conway, & Guey, 1982; Adamatzky, 2010; Johnston & Greene, 2021). The most common statement of the GoL update rules is

1. Loneliness: Any live cell with fewer than two live neighbors dies
2. Survival: Any live cell with two or three neighbors lives to the next generation
3. Overpopulation: Any live cell with more than three live neighbors dies
4. Birth: Any dead cell with exactly three live neighbors comes alive

This set of rules can be stated more compactly as: "A cell will be ON in the next generation if either it is surrounded by three ON cells or it is currently ON and surrounded by exactly two ON cells, otherwise that cell will be OFF", which can be written as

$$\phi(s, \Sigma) = \delta_{\Sigma,3} + s\delta_{\Sigma,2}$$

where $s$ is the current state of that cell, $\Sigma$ is the number of its neighbors that are currently ON, and $\delta_{i,j}$ is the Kronecker delta function, which takes on the value 1 when $i = j$ and 0 otherwise. Alternate formulations also exist, such as (Adachi, Peper & Lee, 2004)

$$\phi(s, \Sigma) = H(2 - (s + 2\Sigma - 6)^2)$$

where $H(\cdot)$ is the Heaviside function, which takes on the value 1 when its argument is greater than 0 and otherwise returns 0.

The global GoL evolution operator $f: \mathcal{G} \to \mathcal{G}$ is then obtained by applying $\phi$ element-wise to every cell in a grid $\mathcal{G}$. A configuration $C$ of an $n \times n$ region of $\mathcal{G}$ is a specific assignment of 0s and 1s to the grid cells in that region. The $(n + 2) \times (n + 2)$ preimage of $C$, denoted $C' = f^{-1}(C)$, is defined as the set $\{C' | f(C') = C\}$. Due to the order-8 $D_4$ symmetry of GoL, the same configuration can appear in multiple ways depending on how it is embedded in the grid. We will refer to these different embeddings of the same configuration as *presentations*.

One of the simplest possible questions we can ask about the preimages of a configuration is Are there any? This is the Garden of Eden (GOE) problem for cellular automata. A configuration $C$ is a Garden of Eden configuration if it has no precursors, that is, if $|f^{-1}(C)| = 0$. A (possibly non-square) minimal subset of a GOE configuration is called an *orphan*. The idea of a GOE configuration derives from the follow theorem:

**The Garden of Eden Theorem** (Moore, 1963; Myhill, 1963): For a $d$-dimensional cellular automaton $f: \Sigma^{\mathbb{Z}^d} \to \Sigma^{\mathbb{Z}^d}$ over a finite alphabet $\Sigma$, $f$ pre-injective $\Leftrightarrow$ $f$ surjective.

The content of this theorem is most easily understood in its contrapositive form, $\neg(f$ pre-injective$) \Leftrightarrow \neg (f$ surjective$)$, which can be written as

$$\exists_{c_1 \neq c_2} f(c_1) = f(c_2) \Leftrightarrow \exists_c |f^{-1}(c)| = 0.$$

In words, a function $f$ fails to be pre-injective if there exist two distinct finite configurations of the same size that are mapped to the same configuration by $f$. Such pairs of configurations are called *twins*. A function $f$ fails to be surjective if there exist configurations without precursors, i.e., GOE configurations. Thus, the theorem states that a cellular automaton possesses GOEs if and only if it has twins. The applicability of this theorem to the Game of Life was noted very early by Alvy Ray Smith (Gardner, 1983:230,248; Poundstone, 1985:49). A simple pair of twins in GoL are the $5 \times 5$ configuration consisting of all OFF cells and the $5 \times 5$ configuration consisting of a single ON cell at its center, both of which evolve to the quiescent state.

Because an $n \times n$ grid in GoL has $2^{(n+2)^2}$ possible precursors, the problem of searching for GOEs is exponential in the grid size. A standard approach is to work backward from a given target configuration, performing a depth-first search through the possible inversions of each cell state filtered by local consistency requirements and backtracking whenever a conflict arises. For the GOE problem, the search can be terminated as soon as the first precursor is found. Hardouin-Duparc (1974) improved on this method by deriving a finite automata that accepted GOEs. More recently, GOE search has been formulated as a satisfiability problem and tackled with SAT solvers (Bain, 2007; Hartman et al., 2013; Knuth, 2015).

## 3. A Map of the Territory

The current status of our ongoing GOE search is shown in Figure 1 as a function of grid density $\rho$ and grid size $n$. From previous work, it was known that no GOEs exist for grid sizes of $6 \times 6$ or smaller, that GOEs exist for grid sizes of $10 \times 10$ and higher, and that no GOEs satisfying certain symmetries exist at intermediate grid sizes (Hartman et al., 2013). For the results in this section, we employ a very efficient iterative implementation of backward search with all consistency-checking reduced to table lookups. In addition, we take full advantage of the $D_4$ symmetry of the Game of Life in order to reduce the number of unique configurations that must be examined. Finally, since our GOE searches are organized by density, grid configurations of a given density are generated in cool lex order, which has a very fast branch-free implementation (Ruskey & Williams, 2009). Using this approach, we have extended previous results in three ways.

[Insert Figure 1 Here]

First, we have shown that no GOEs exist in $7 \times 7$ grids, a result that took several years to obtain given that there are 562 949 953 421 312 configurations to be examined.

Second, we have shown that no GOEs exist outside an $n$-dependent range of densities for $8 \leq n \leq 11$ (red region). Since the number of $n \times n$ grids containing $k$ 1s is $\binom{n^2}{k}$, this region represents the elimination of an additional 271 687 723 328 802 configurations, for a total of 834 706 429 847 348 lying in the red region of the plot. This result also took several years to obtain. Note that the number of possible configurations of an $n \times n$ grid containing between $l$ and $m$ ON cells is given by

$$\sum_{i=l}^{m} \binom{n}{i} = \binom{n}{l} {}_2F_1(1, l-n; l+1; -1) - \binom{n}{m+1} {}_2F_1(1, m-n+1; m-n; -1)$$

where ${}_2F_1$ is the ordinary hypergeometric function.

Third, we have extended the number of GOEs which are known to exist and the densities over which they exist for $n = 10, 11$ (blue region), providing a large catalogue of GOEs for analysis. These GOE configurations will be examined in more detail in Sections 5 and 6.

The white region in Figure 1 indicates the grid sizes and densities that remain to be systematically examined, representing a total of $10^{83}$ configurations. Given the amount of computation that would be required to complete this map, it is worth exploring other perspectives that might provide important insights into the nature and layout of GOEs in this space.

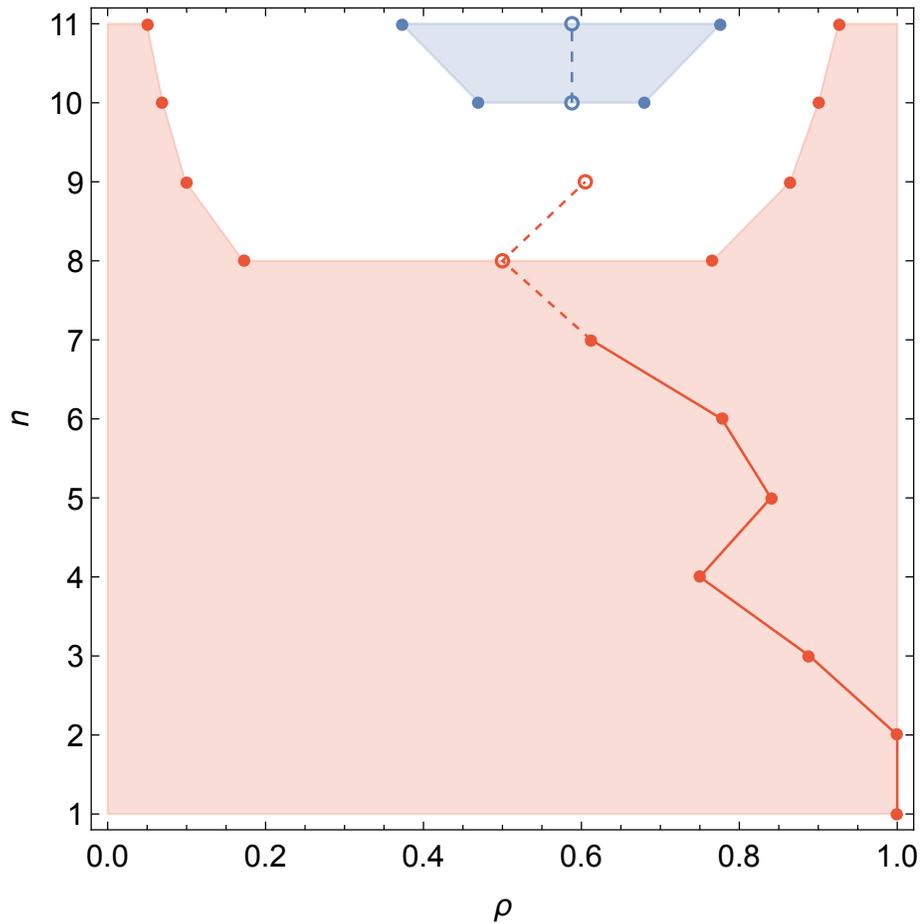

**Figure 1:** Current state of knowledge regarding GOE configurations in GoL for square grids of size $n$ and density $\rho$. Regions where GOE configurations are known to exist are shown in blue. Regions where GOE configurations have been proven to *not* exist are shown in red, with all results above $n = 6$ new to this paper. The white gap between these two regions indicates the current frontier where the question is not yet settled. The numbers of ON cells bounding the white region are as follows: $8 \times 8 : [12,48]$ , $9 \times 9 : [9,69]$ , $10 \times 10 : [8,89]$ , $11 \times 11 : [7,111]$. The density of the configuration with the minimum number of precursors for each $n$ is marked with red points connected by solid red lines, with open red circles connected by dashed red lines indicating incomplete computations that are still in progress. Finally, the current mean density at which GOE configurations occur is marked with open blue circles connected by a dashed blue line.

## 4. Approaching the Garden

Our first strategy is to embed the GOE problem in a more general problem. If a GOE is a configuration with no precursors, then perhaps we should examine how the number of precursors that a configuration can have varies with grid size and density. In particular, we will focus on those configurations with the smallest number of precursors as we consider larger and larger grids. Unfortunately, because we cannot know a priori which configurations will have the smallest number of precursors, we must compute the number of precursors for all configurations of a given grid size and density in order to find the smallest. But even though precursor counting is more computationally expensive than GOE testing, it carries more structure than a simple binary classification.

Our basic object of study in this section is a *precursor count envelope* (Figure 2). For a given grid size, this is simply a log plot of the minimum and maximum number of precursors as a function of density. The mean number of precursors is also indicated. To date, precursor count envelopes for up to $n = 6$ have been completed using the incremental aggregation algorithm with all applicable optimizations (Beer, 2017). The $n = 7$ envelope is currently in progress ($2^{49}$ total configurations; 71.67% completed).

[Insert Figure 2 Here]

Precursor count envelopes have a characteristic "triangular" shape, which becomes more defined as $n$ increases. The leftmost vertex, which corresponds to the vacuum state (all 0s), always has the most precursors, whereas the rightmost vertex, which corresponds to all 1s, does not generally have the fewest precursors. The number of precursors of these two configurations grows as $O(2^{n^2})$ and $O(2^n)$, respectively (Beer, 2017). Between these two extremes we see a relatively straight left edge broken by a single kink, a long diagonal right edge punctuated by small peaks, and an increasingly flat bottom edge with a single minimum peak. Note that the apparent flattening of the edges and peaks is at least partly due to the logarithmic vertical scaling of these plots.

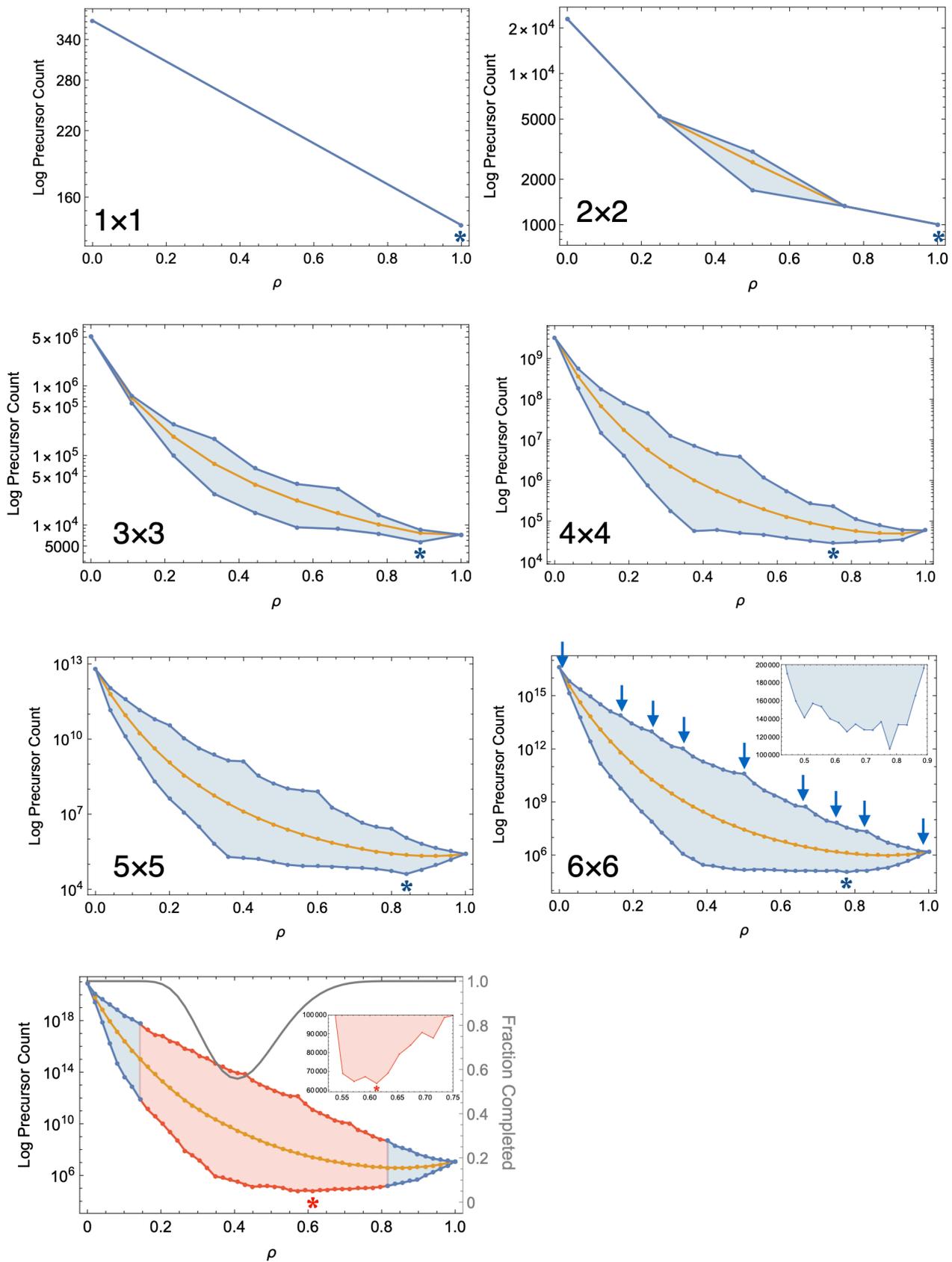

**Figure 2:** Precursor count envelopes for grids up to size $7 \times 7$. The upper and lower curves in each plot bound give the log precursor count ranges for each density, while the orange central orange denotes the mean log precursor count. Asterisks mark the location of global minima, whereas the arrows for the $6 \times 6$ case indicate the local maxima referred to in the main text. Non-log closeups of the minima are shown as insets for the $6 \times 6$ and $7 \times 7$ precursor envelopes. Density ranges for which the search is complete are shown in blue, while density ranges for which the search is ongoing are shown in red. The state of completion for the incomplete density region of the $7 \times 7$ precursor envelope is shown as a gray curve whose scale is indicated on the right.

Let us begin with the global minima along the bottom edges of the precursor count envelopes (indicated with asterisks in Figure 2). These represent the configuration(s) that are closest to being GOEs for each grid size because they have the smallest possible number of precursors for that size. These minimal configurations are shown in Figure 3A, along with their precursor counts. A few interesting observations can be made. First, whereas the $n = 5$ minimal configuration clearly builds on the $n = 3$ case, the $n = 4, 6$, and 7 minimal configurations are each unique and become increasingly intricate as grid size increases. Second, the minimal configurations possess very different symmetries, with symmetry group orders of 1, 2, or 4. Third, the minimal precursor count grows with grid size up to $n = 6$, but then *decreases* for $n = 7$. We can see this pattern even more clearly if we plot maximal and minimal precursors counts for $n \leq 10$ (Figure 3B). This plot shows that, while the maximum curve continues to grow with $n$, the minimum curve turns around after $n = 6$, eventually plunging to $-\infty$ in this log plot for $10 \times 10$ grids, which are known to contain GOEs. A final observation is that the densities at which the minimal configurations occur exhibits a meandering but generally decreasing trend as grid size increases (red lines in Figure 1)

[Insert Figure 3 Here]

Although not directly related to our study of GOE configurations, another feature of these precursor count envelopes is worth exploring. As mentioned above, the upper limits of precursor count envelopes exhibit a series of peaks at regular intervals (indicated by arrows for the $6 \times 6$ case in Figure 2). The number of peaks increases with grid size and they also become narrower and sharper. In addition, the largest peaks occur at intermediate densities and then decay in size as we move toward the extremes. A closer examination reveals that, as the maximum precursor count decreases with increasing density, each "peak" actually consists of a flattening out of the curve followed by a sharp drop. Interestingly, the peaks themselves occur at densities which are specific fractions of $n^2$. For example, for the $6 \times 6$ case, we see peaks at $\rho = 0, 1/6, 1/4, 1/3, 1/2, 2/3, 3/4, 5/6$, and 1. Although the $7 \times 7$ envelope exhibits similar features, we can make no detailed quantitative comparison yet since this envelope is still under construction.

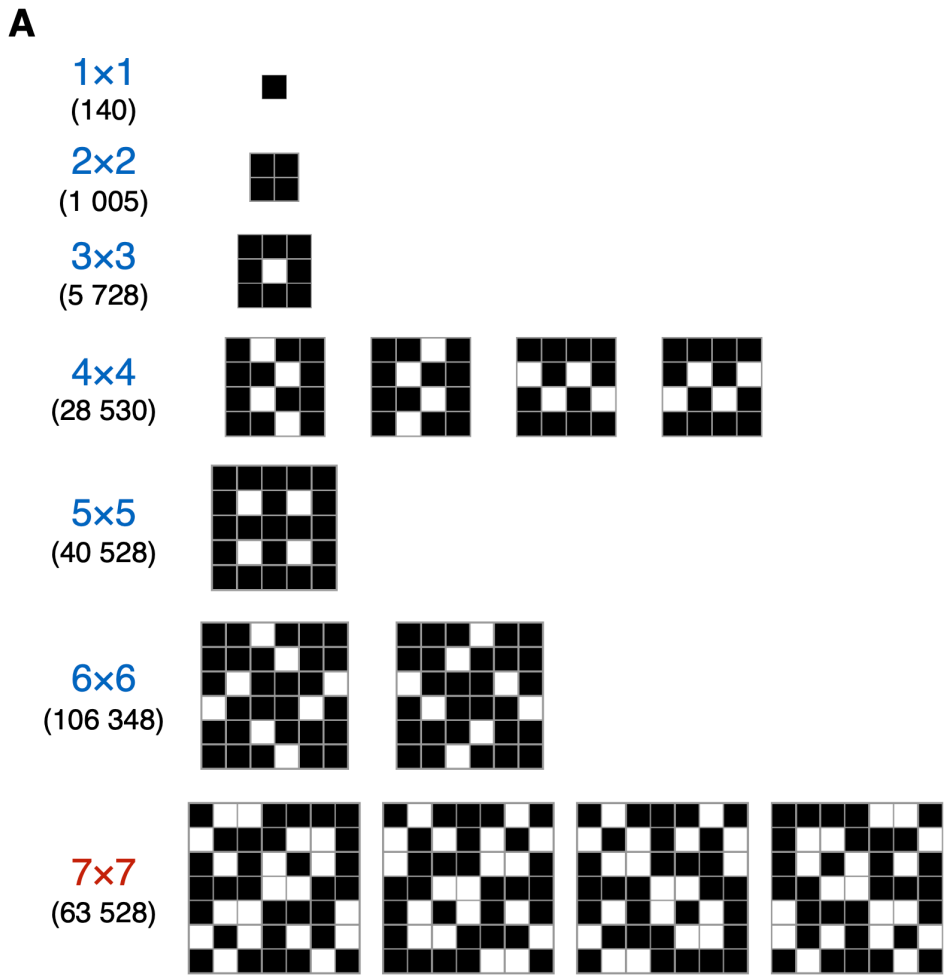

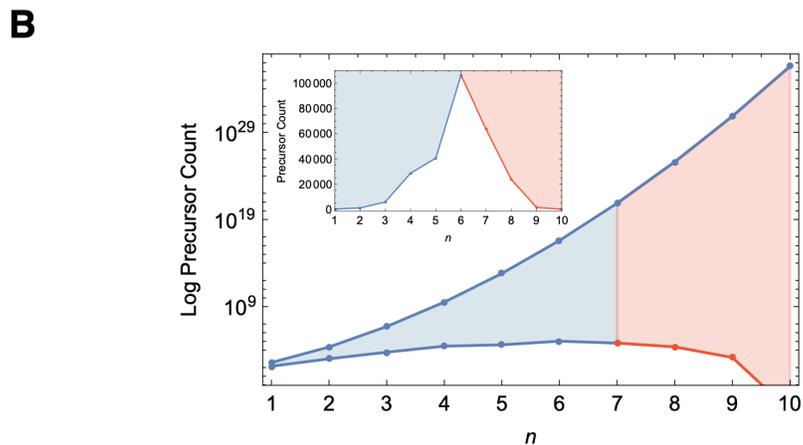

**Figure 3:** The progression of minimal precursor counts (asterisks in Figure 2) with increasing grid size. (A) The configurations with the fewest number of precursors for each grid size up to $7 \times 7$. ON cells are displayed in black and OFF cells are displayed in white. The number of precursors for each configuration is given in parentheses. The minimum precursor is always unique, but may have multiple presentations due to its symmetries. Note that the $7 \times 7$ (red) computation is still ongoing, so that it is possible that a configuration with a smaller number of precursors might still be found for this case. (B) Precursor count ranges for grids up to size $10 \times 10$. The inset is a non-log plot of the minimum curve of the main log plot, more clearly showing the sharp turnaround at $n = 6$. Here blue curves indicate completed computations and red curves indicate ongoing computations. However, it is important to emphasize that even the red portions constitute *upper bounds* for the minimum curve; additional data can only *lower* this portion of the curve if configurations with fewer precursors are subsequently found.

What underlies these patterns in the maximum precursor counts for a given grid size? As Figure 4A illustrates for $6 \times 6$ grids, these peaks occur for configurations that are particularly regular, such as stripes or blocks cells in the same state. As a general rule, contiguous empty regions have the largest number of precursors and contiguous filled regions have the smallest number. Perhaps the peak configurations represent a balance between these two effects. One has the impression that, as more ON cells are added, the precursor count asymptotes until some packing limit is reached and then a reorganization occurs into a new pattern. An examination of a representative maximal configuration for each possible number of ON cells shows that this is indeed the case (Figure 4A). A similar sequence of reorganizations is observed for minimal precursor count configurations, but here the patterns are much more complicated (Figure 4B).

[Insert Figure 4 Here]

## 5. Cataloguing the Garden

Our second strategy is to expand the set of known $10 \times 10$ and $11 \times 11$ GOEs as much as possible in order to examine their collective properties. We will start by reconstructing the symmetric GOEs identified by Hartmann, Heule, Kwekkeboom and Noels (henceforth HHKN) (Hartman et al., 2013) and then build from there. We will make use of the only two symmetry operators defined by HHKN for a square grid $\mathcal{G}$ that produce $10 \times 10$ and $11 \times 11$ GOEs:

$\sigma_{90}$ (90 degree rotation): $\mathcal{G}_{i,j} = \mathcal{G}_{j,n+1-i}$ for $i,j \in \{1, \cdots, n\}$

$\{\sigma_D, \sigma_{180}\}$ (combined diagonal reflection and 180 degree rotation): $\mathcal{G}_{i,j} = \mathcal{G}_{j,i} = \mathcal{G}_{n+1-i,n+1-j}$ for $i,j \in \{1, \cdots, n\}$

For the $10 \times 10$ case, we found 8 GOEs satisfying $\sigma_{90}$ and 62 GOEs satisfying $\{\sigma_D, \sigma_{180}\}$, for a total of 70. These are identical to the values reported by HHKN. We then generated all 1-, 2-, 3-, 4-, and 5-bit perturbations from this set of symmetric GOEs, obtaining 456, 1740, 4864, 10 384 and 17 036 new GOEs, respectively. Finally, we discovered an additional 1000 GOEs through less systematic random sampling, for a grand total of 35 550. The density histogram of this sample of $10 \times 10$ GOEs is shown in Figure 5A. It is roughly normal with a mean of 0.588 and a standard deviation of 0.0295, but exhibits significant fluctuations. The minimum number of ON cells in this set of GOEs is 47 (Figure 5B, left) and the maximum is 68 (Figure 5B, right). Removing D4-symmetric redundancy from our full set of $10 \times 10$ GOEs reduces their total to 4 617.

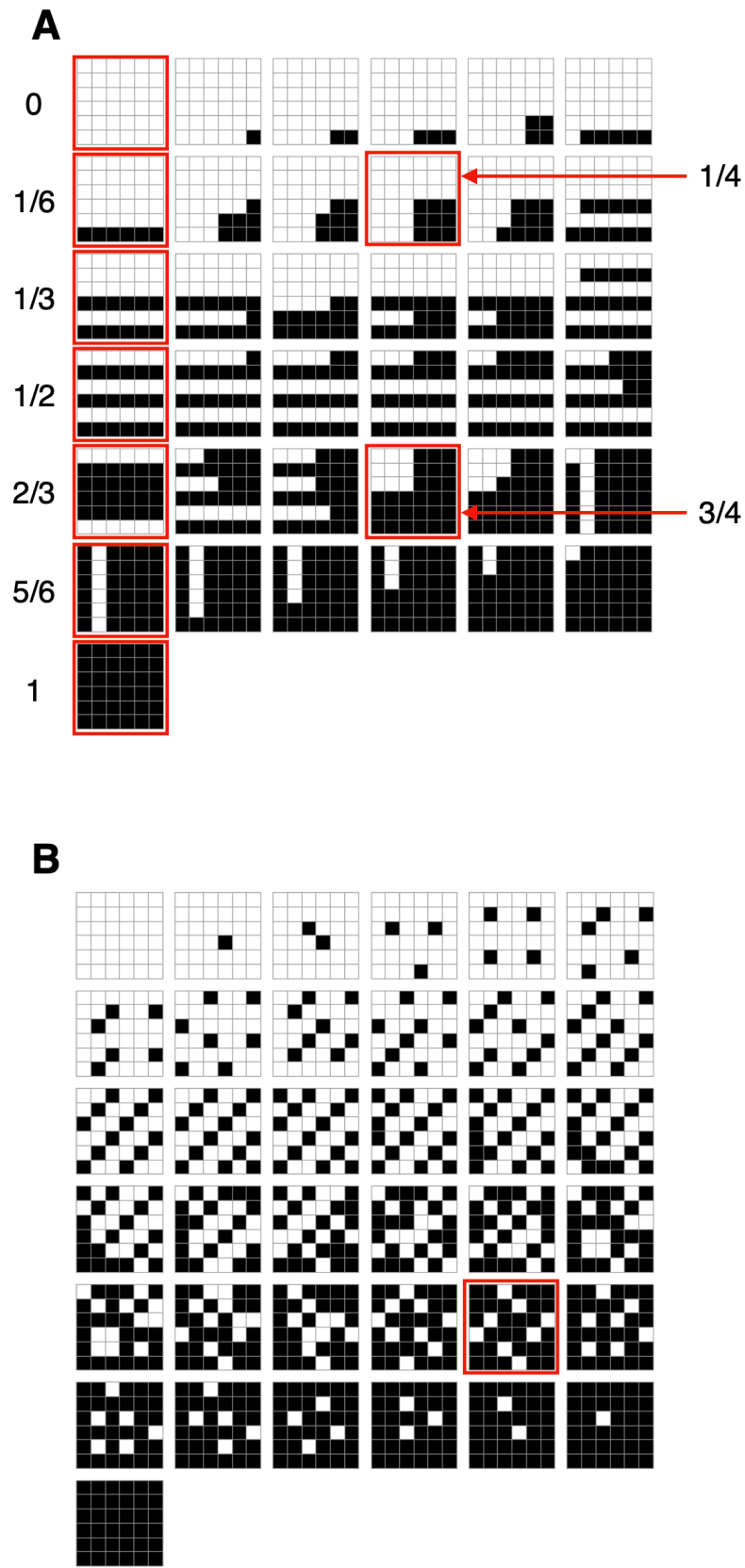

**Figure 4:** A representative example of each extremal precursor count configuration in $6 \times 6$ grids as the number of ON cells varies from 0 to 36. ON cells are displayed in black and OFF cells are displayed in white. (A) Maximal precursor count configurations. The configurations corresponding to the maximal precursor count peaks in the $6 \times 6$ envelope (denoted by arrows in Figure 2) are outlined in red along with their corresponding densities. (B) Minimal precursor count configurations. The globally minimal configuration is outlined in red.

[Insert Figure 5 Here]

For the $11 \times 11$ case, we found 41 702 GOEs satisfying $\sigma_{90}$. This is 342 fewer than the 42 044 reported by HHKN. However, it is interesting to note that removing D4-symmetric redundancy from our set leaves us with 21 022 GOEs, which is exactly half of HHKN's reported number. We then generated all 1-bit perturbations of this set of symmetric GOEs, obtaining 565 280 new GOEs for a grand total of 606 982. The density histogram of this sample of $11 \times 11$ GOEs is shown in Figure 6A. It is roughly normal with a mean of 0.588 and a standard deviation of 0.0025, but exhibits significant fluctuations. The minimum number of ON cells in this set of GOEs is 45 (Figure 6B, left) and the maximum is 94 (Figure 6B, right). Removing D4-symmetric redundancy from our full set of $11 \times 11$ GOEs reduces their total to 91 900.

[Insert Figure 6 Here]

What are the prospects for finding any $8 \times 8$ or $9 \times 9$ GOE configurations? HHKN showed that no symmetric $8 \times 8$ or $9 \times 9$ GOEs exist, but the nonsymmetric cases remain open. The only evidence we have for this question comes from Figure 1, where we observe that the (currently incomplete) boundaries of the blue GOE region, as well as its current mean GOE 1-densities, all seem to be converging to a point around $n = 9$. *If* these boundaries are close to the final ones, and *if* they can be linearly extrapolated, then this suggests that there will be no $8 \times 8$ GOEs, but that $9 \times 9$ GOEs might exist in a narrow region around a density of $\rho = 0.59$ (corresponding to about 47 ON cells). Interestingly, the $9 \times 9$ configuration with the fewest precursors (1394) that we have so far observed contains 49 ON cells.

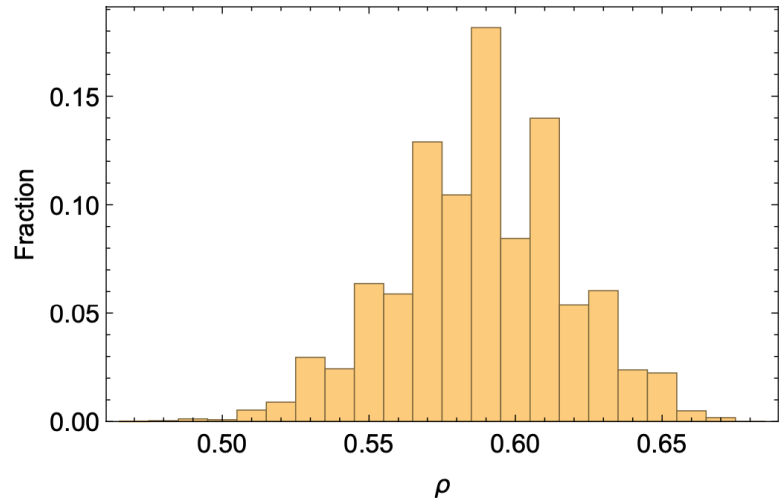

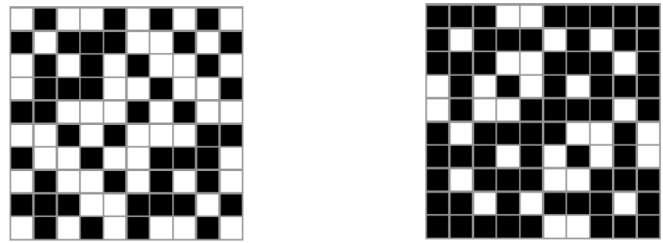

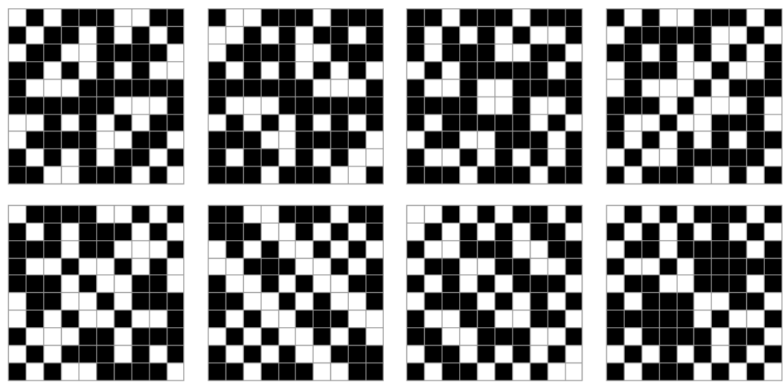

**Figure 5:** Properties of our set of 35 550 $10 \times 10$ GOEs. (A) The density histogram for this set, with mean 0.588 and standard deviation 0.0295. (B) Presentations of the $10 \times 10$ GOE configuration with the least (47; left) and most (68; right) ON cells. (C) A sample of some of the other symmetric GOEs from our set.

A

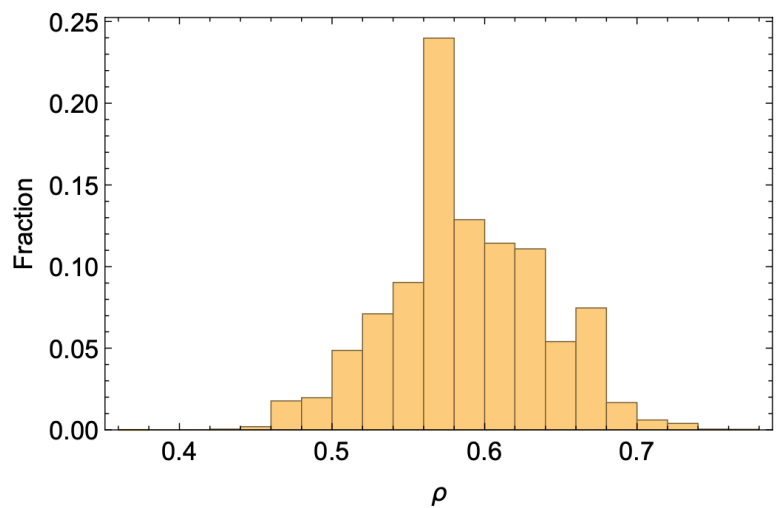

B

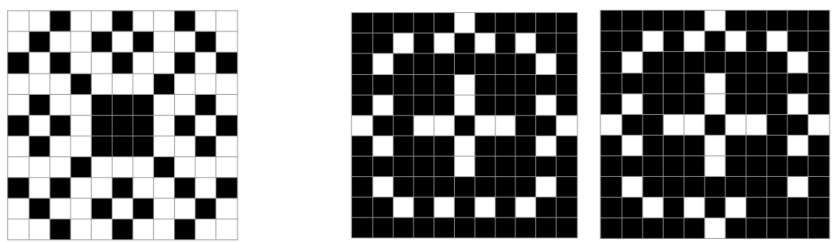

C

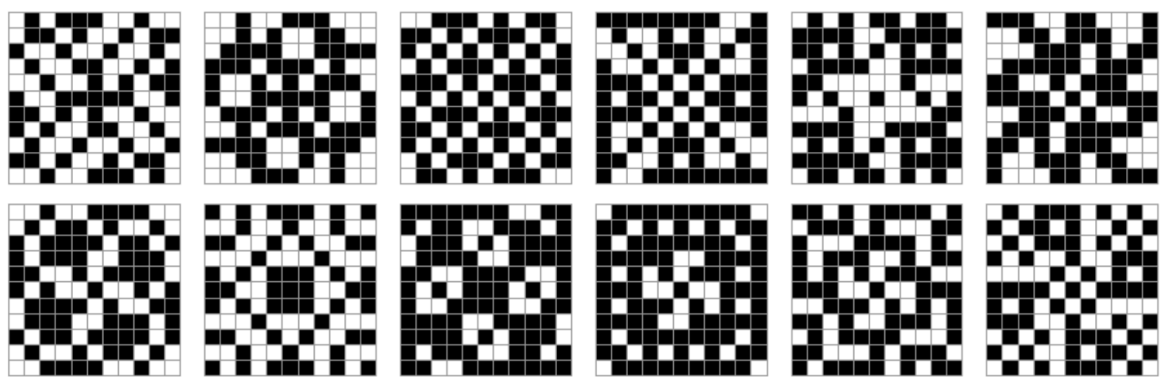

**Figure 6:** Properties of our set of 606 982 11 × 11 GOEs. (A) The density histogram for this set, with mean 0.588 and standard deviation 0.0025. (B) Presentations of the 11 × 11 GOE configuration(s) with the least (45; left) and most (94; right) ON cells. (C) A sample of some of the other symmetric GOEs from our set.

## 6. The Anatomy of Flowers of Eden

Our third and final strategy is to begin to examine the structure of the networks of constraints that underlie GOEs. A first step in this direction is to consider the following *dilation decomposition* of a GOE. Beginning with the central set of cells in an $n \times n$ GOE (a single cell if $n$ is odd and a $2 \times 2$ grid if $n$ is even), we count the number of precursors that each subsequent increase in window size by n+2 as we incrementally dilate our window into the GOE until its full extent is reached (which has 0 precursors by definition). Our goal is to examine the scaling of the precursor count as we approach the GOE. An example of this decomposition for a $10 \times 10$ GOE is shown in Figure 7A. Figures 7B and 7C compare this scaling for GOEs (blue) and density-matched nonGOEs (red) in $10 \times 10$ and $11 \times 11$ grids, respectively.

[Insert Figure 7]

Several interesting observations can be made about these plots. For small window sizes, GOEs and nonGOEs are indistinguishable. In fact, on average, the central cores of GOEs actually have more precursors than those of nonGOEs in the $10 \times 10$ case. Initially, both sets of curves also grow exponentially with increasing window size. However, because the rates of growth are smaller for the GOE curves, these curves quickly approach the lower limit of the range of precursor counts. Eventually, the GOE curves turn around completely, actually decreasing with increasing window size before plunging to $-\infty$ in this log plot. Thus, we can clearly see the interplay between two factors: the exponential growth in precursor count with increasing grid size and the exponential decay of precursor count with increasing constraints. Although the network of possible constraints on precursors is initially outpaced by the exponential growth of the number of possible precursors with grid size, it eventually tightens sufficiently for $n \geq 6$ for GOEs to occur.

Note that the increasingly wide precursor count separation between the dilation decomposition curves for GOEs and nonGOEs does not mean that near misses do not occur. For example, if we flip just one bit of the $10 \times 10$ GOE shown in Figure 7A, we can find nonGOEs with as few as 14 precursors. Also note that analogous precursor count scaling curves can be constructed for any other incremental approach to a GOE, such as the greedy spiral method that Nicolay Beluchenko has used to construct a GOE (Sloane, 2011).

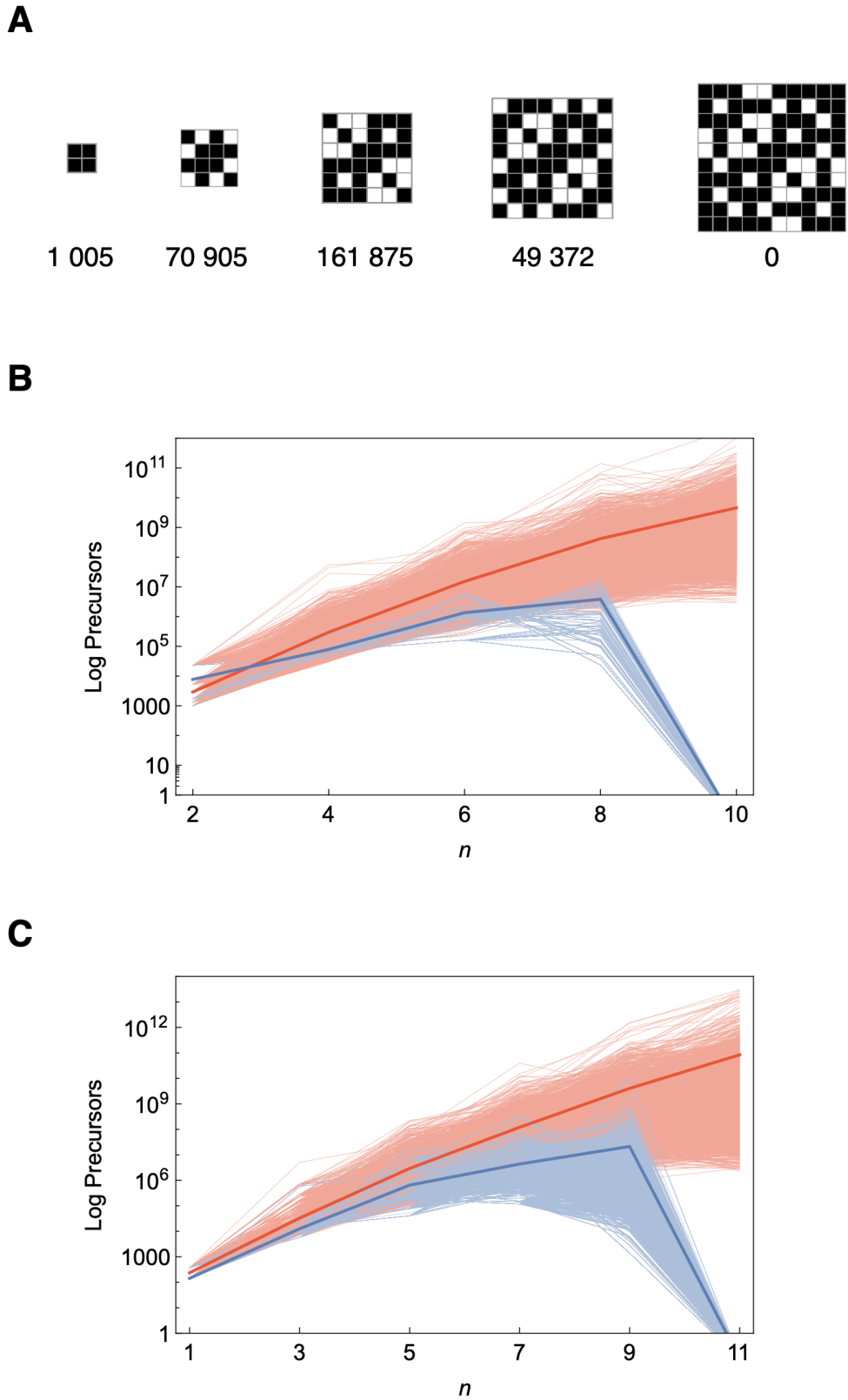

**Figure 7:** Precursor scaling across dilation decompositions of GOE (blue) and density-matched nonGOE (red) configurations. (A) A dilation decomposition of the $10 \times 10$ GOE on the right, with the number of precursors shown below each step of the decomposition. (B) Precursor scaling across dilation decompositions of our D4-symmetry-reduced set of 4 617 $10 \times 10$ GOEs (blue) compared to the same number of density-matched nonGOEs (red), with individual curves shown with lighter thin lines and means shown with darker thick lines. (C) Precursor scaling across dilation decompositions of a random sample of 5 000 configurations from our D4-symmetry-reduced set of 91 900 $11 \times 11$ GOEs compared to the same number of density-matched nonGEOs.

These results highlight the special nature of the network of constraints that underlie GOEs. It would thus be very interesting to attempt to characterize such networks in more detail. While such a project lies outside the scope of the current paper, it is worth briefly visualizing how it might work. For example, consider the precursors of the {ON, OFF} configuration consisting of an ON cell adjacent to an OFF cell (Figure 8). Imagine stacking the 140 possible $3 \times 3$ precursors for an ON cell and the 372 possible precursors of an OFF cell (Figure 8, left). This generates up to $140 \times 372 = 52\,824$ possible precursor configurations for the joint {ON, OFF} pattern.

[Insert Figure 8]

However, because these two cells are adjacent in the grid, the right two columns of a potential ON precursor and the left two columns of a potential OFF precursor (shown in red) must match for any pairing that forms a valid precursor to the joint configuration (Figure 8, right). This determines a constraint network (red lines), or, equivalently, a consistency network (green lines) that strongly restricts the allowable pairings. If we incrementally add one component at a time to analyze larger and larger patterns, then these networks will always be (bipartite) graphs. However, if we consider adding multiple components of a target pattern simultaneously, then both the constraint and consistency relations in general define hypergraphs. In either case, in this formulation GOEs correspond to networks whose consistency set is empty. Note that this representation makes explicit both the growth of *possible* pairings due to the multiplicative character of their combination and the decay of *allowable* pairings due to the increasing numbers of constraints that are brought to bear on these combinations.

## 7. Conclusion

In this paper, we have extended Hartman et al.'s (2013) systematic study of Garden of Eden configurations in the Game of Life in four ways. First, we have constructed a map of the currently known layout of GOEs and nonGOEs in density/size space. The most notable result here was proving that no $7 \times 7$ GOE configurations exist, but we also tightened the bounds over which $10 \times 10$ and $11 \times 11$ GOEs can exist. Second, we have used precursor count envelopes to characterize how the density-dependent structure of the number of precursors varies with increasing grid size as we approach the point where GOEs begin to occur. The significant result here was showing that the minimal precursor count, which initially grows with grid size, turns around and begins to shrink after $n = 6$, but we also identified the minimal precursor count configurations up to $7 \times 7$ and studied features of the maximal precursor count configurations as well. Third, we expanded the set of known GOEs up to a grid size of $11 \times 11$. Finally, we initiated a study of the internal structure of GOEs by introducing dilation decomposition of GOEs, studying the precursor count scaling of such decompositions, and sketching a way to visualize the networks of constraints that underlie GOEs.

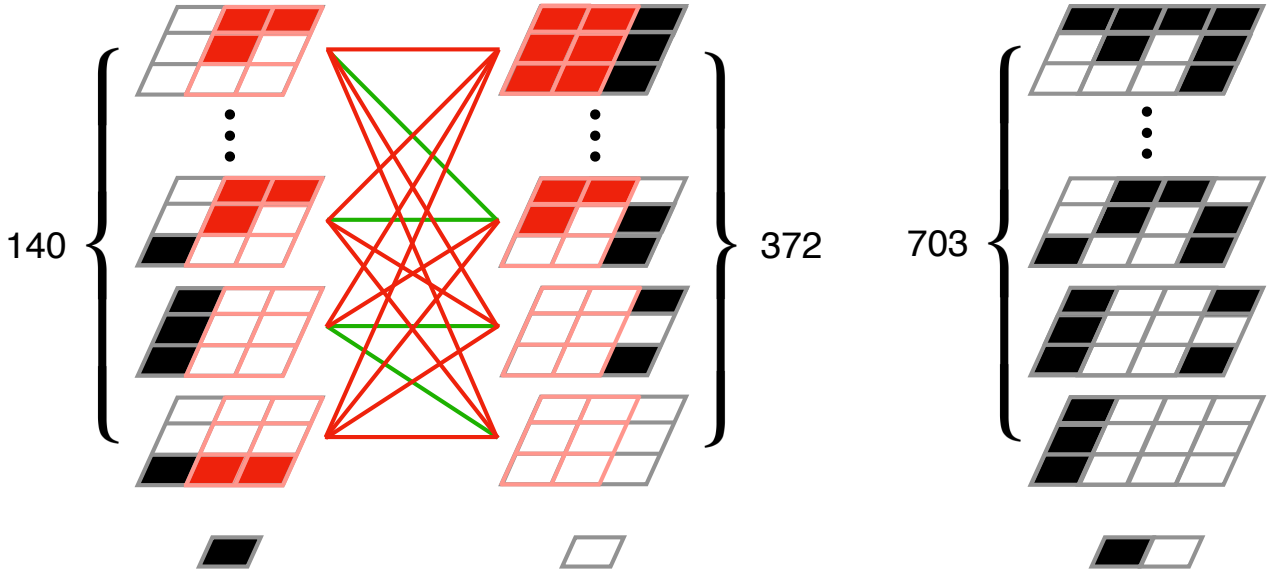

**Figure 8:** Visualization of the constraint (red) or, equivalently, consistency (green) network for the precursors of a simple $2 \times 1$ configuration consisting of an ON adjacent to an OFF cell. (Left) The possible $3 \times 3$ precursors for each target cell individually, with the overlapping portions that constrain their combination highlighted in red. (Right) An illustration of the final set of $4 \times 2$ precursors of the joint {ON, OFF} target configuration that satisfy the consistency/constraint network.

There are a number of directions for future work. First, the analysis in this paper is still woefully incomplete. The density/size map, the precursor count envelopes, and the set of known GOEs all require significant expansion, with a special focus on the $9 \times 9$ and $9 \times 9$ grids with densities in the range of 0.55-0.65 where their minimal precursor count configurations seem to reside. This will obviously require significantly more computational resources. In the long run, it is likely that SAT approaches are better for GOE searches, but something like the incremental aggregation algorithm may still be well-suited for precursor counting. Second, it would be interesting to extend the analysis to nonsquare grids, with the current density/size map representing merely a slice through this now 3-dimensional space. Third, it would be useful for the community to set up and maintain a public database of known GOE configurations, ideally with an associated orphan database. Finally, and perhaps most fundamentally, a serious study of the structure of the networks of constraints underlying GOE vs. nonGOE configurations needs to be undertaken.

*Acknowledgments*: Computation performed for this research was supported in part by Lilly Endowment, Inc., through its support for the Indiana University Pervasive Technology Institute, and in part by the Indiana METACyt Initiative. The Indiana METACyt Initiative at IU is also supported in part by Lilly Endowment, Inc.

# References

Adachi, S., Peper, F., and Lee, J. (2004). The Game of Life at finite temperature. *Physica D* **198**:182-196.

Adamatzky, A. (Ed.) (2010). *Game of Life Cellular Automata*. Springer.

Bain, S. (2007). Time-reversal in Conway's Life as SAT. In *Proceedings of the 20th Australian joint conference on Advances in artificial intelligence* (pp. 614–618). Springer-Verlag.

Beer, R.D. (2017). Computing aggregate properties of preimages for 2D cellular automata. *Chaos* **27**: 111104.

Berlekamp, E.R., Conway, J.H. and Guy, R.K. (1982). *Winning Ways for Your Mathematical Plays, Vol. 2*. A. K. Peters.

Flammenkamp, A. (2009). Garden of Eden / Orphan. http://wwwhomes.uni-bielefeld.de/achim/orphan.html. Retrieved on October 10, 2022.

Gardner, M. (1970). The fantastic combinations of John Conway's new solitaire game "life". *Scientific American* **223**(4):120-123.

Gardner, M. (1983). *Wheels, Life, and Other Mathematical Amusements*. W.H. Freeman.

Hardouin-Duparc, J. (1974). Paradis terrestre dans l'automate cellulaire de Conway. *Rev. Française Automat. Informat. Recherche Operationnelle Ser. Rouge* **8**:64–71.

Hartman, C., Heule, M.J.H., Kwekkeboom, K. and Noels, A. (2013). Symmetry in gardens of eden. *The Electronic Journal of Combinatorics* **20**(3):P16 https://doi.org/10.37236/2611

Johnston, N. and Greene, D. (2021). *Conway's Game of Life: Mathematics and Construction*. Lulu.com.

Knuth, D. (2015). *The Art of Computer Programming, Volume 4, Fascicle 6: Satisfiability*. Addison-Wesley.

LifeWiki (2009). "Garden of Eden" (https://conwaylife.com/wiki/Garden_of_Eden) Retrieved on June 6, 2022.

Moore, E.F. (1963). Machine models of self-reproduction. *Proc. Symp. Appl. Math., Vol 14* (pp. 17-34). American Mathematical Society.

Myhill, J. (1963). The converse of Moore's Garden-of-Eden theorem. *Proc. Amer. Math. Soc. Vol 14* (pp. 685-686). American Mathematical Society.

Poundstone, W. (1985). *The Recursive Universe*. William Morrow and Company.

Ruskey, F. and Williams, A. (2009). The coolest way to generate combinations. *Discrete Mathematics* **309**(17):5305-5320.

Sloane, N.J.A. Sequence A196447 in *The On-Line Encyclopedia of Integer Sequences*. oeis.org/A196447. Retrieved August 17, 2022.